\documentclass[
  aps,
  prd,
  11pt,
  superscriptaddress,
  notitlepage,
  onecolumn,
  amsmath,
  nofootinbib
]{revtex4-1}

\AtBeginDocument{
  \linespread{1.05}\selectfont
}

\usepackage[
  a4paper,
  left=2.3cm,
  right=2.3cm,
  top=2.5cm,
  bottom=2.5cm
]{geometry}

\usepackage{microtype} 
\usepackage{graphicx}
\usepackage{xcolor}
\usepackage{orcidlink}
\usepackage{enumerate}
\usepackage{enumitem}
\usepackage{ulem}
\definecolor{linkcolor}{rgb}{0.0,0.3,0.5}
\definecolor{linkcolor}{rgb}{0.0,0.3,0.5}
\definecolor{mypurple}{RGB}{143, 116, 210}

\graphicspath{{Figures/}}

\newcommand{\hias}{School of Fundamental Physics and Mathematical Sciences, Hangzhou Institute for Advanced Study, University of Chinese Academy of Sciences, Hangzhou 310024, China}

\begin{document}

\title{Vanishing Island: Spectral Criticality in Boson Stars}

\author{Bo-Xuan Ge
\orcidlink{0000-0003-0738-3473}}
\email{bo-xuan.ge@ucas.ac.cn}
\affiliation{\hias}

\begin{abstract}
We identify a cusp-type endpoint at which a finite radially stable
region of solitonic boson stars disappears.  Two fundamental
radial zero-mode boundaries coalesce there, while the constrained
quadratic stiffness and cubic coefficient vanish and the quartic
coefficient remains finite and non-zero.  The soft radial mode obeys
\(\Omega_{\mathrm{soft}}\propto
(\sigma_{\mathrm{cusp}}-\sigma)^{1/2}\), revealing a divergent
linear timescale at the endpoint.  The same critical structure
persists under the tested deformation of the scalar self-interaction.
\end{abstract}

\maketitle

\section{Introduction}

Self-gravitating bosonic fields admit families of compact equilibrium
configurations whose radial stability can change as the central field
amplitude or the self-interaction is varied
\cite{Kaup:1968zz,Ruffini:1969qy,Gleiser:1988rq,Kusmartsev:1990cr,
Alcubierre:2003sx,Liebling:2012fv,Ge:2024itl,Ge:2025btw}.
Solitonic boson stars provide a particularly clear example
\cite{Friedberg:1986tq,Kleihaus:2011sx,Marks:2025jit,Marks:2026yle}.
Previous nonlinear evolutions identified a restabilised branch, denoted
region III in Ref.~\cite{Ge:2024itl}, lying between two unstable
branches.  When \(\sigma\) is treated as a second control parameter,
this stable interval extends into a finite region of the
\((|\phi_c|,\sigma)\) family.  We show here that, as \(\sigma\) is increased, the surrounding
unstable region closes in on the stable island, gradually engulfing
it until the island \textit{vanishes} at the coalescence of its two
radial-stability boundaries.

Turning-point and catastrophe arguments provide a useful global
description of equilibrium families
\cite{Kusmartsev:1990cr,Tamaki:2010zz,Kleihaus:2011sx},
but an extremum of an equilibrium
curve does not by itself identify the fundamental radial zero mode or
determine the number of unstable radial modes~\cite{Santos:2024vdm}.  The disappearance of
the stable island therefore raises a sharper question: what happens to
the radial spectrum when its two boundaries merge?  In particular,
cusp-like equilibrium geometry alone is not sufficient to establish a
higher-order critical endpoint; the fundamental radial zero mode and
the leading nonlinear response along the corresponding soft direction
must also be resolved.

A précis of our results is as follows.  We construct the stationary
Einstein--Klein--Gordon family over \((|\phi_c|,\sigma)\) and compute
the radial pulsation spectrum throughout the endpoint region.  The
stable island is bounded by two curves on which the fundamental radial
mode has zero frequency.  These curves approach one another with a
square-root separation and merge at a cusp-type endpoint.  At the
endpoint, the fundamental radial mode becomes soft, while the
constrained quadratic stiffness and cubic coefficient vanish and the
quartic coefficient remains finite and non-zero.

Approaching the endpoint from below, the soft radial frequency obeys
\begin{equation}
    \Omega_{\mathrm{soft}}
    \propto
    (\sigma_{\mathrm{cusp}}-\sigma)^{1/2},
\end{equation}
corresponding to a divergent linear soft-mode timescale. Since the
stable-island width
\(\Delta|\phi_c|\equiv|\phi_c|_+-|\phi_c|_-\)
obeys the same square-root scaling, the two results imply
\(\Omega_{\mathrm{soft}}\propto\Delta|\phi_c|\) near the endpoint.
This links the geometric collapse of the stable island to the
closing of its radial spectral gap. The local fixed-charge
equilibrium structure is that of a non-degenerate cusp, and the
same critical endpoint persists under the tested sextic deformation
of the scalar potential. The vanishing stable island is therefore
characterized not only by the coalescence of equilibrium branches,
but also by the associated softening of the fundamental radial mode
and the higher-order degeneracy of the constrained response.

\section{Model}    

We consider a spherically symmetric complex scalar field with the
stationary ansatz
\begin{equation}
    \phi(t,r)=|\phi(r)|e^{i\omega t},
\end{equation}
where \(|\phi(r)|\) is the radial scalar-field amplitude and
\(\omega\) is the stationary scalar-field frequency.  The spacetime
metric is written as
\begin{equation}
    ds^2
    =
    -e^{2\Phi(r)}dt^2
    +X(r)^2dr^2
    +r^2d\Omega^2 ,
\end{equation}
with \(r\) the areal radius.  We use the solitonic potential~\cite{Friedberg:1986tq,Kleihaus:2011sx}
\begin{equation}
    V(\phi)
    =
    |\phi|^2
    \left(
        1-\frac{2|\phi|^2}{\sigma^2}
    \right)^2 ,
\end{equation}
where \(\sigma\) controls the self-interaction strength.  We work in
units \(G=m=1\), with \(m\) the scalar-field mass.

The stationary solutions form a two-parameter family labelled by
\((|\phi_c|,\sigma)\), where
\(|\phi_c|\equiv|\phi(0)|\) is the central scalar-field amplitude.
For each pair \((|\phi_c|,\sigma)\), we solve the spherically symmetric
Einstein--Klein--Gordon boundary-value problem following
Ref.~\cite{Helfer:2021brt}, obtaining the scalar and metric profiles
together with \(\omega\).  The global \(U(1)\) symmetry of the complex
field gives the conserved Noether charge \(Q\), which will also enter
the constrained description of the endpoint below.

\section{Radial stability}    

We determine radial stability from the linear pulsation spectrum
\cite{Kain:2021rmk}.  Let \(u_0\) denote collectively a stationary
boson-star background.  Linearising about \(u_0\) and separating
radial perturbations with harmonic time dependence
\(e^{i\Omega_n t}\) gives a coupled radial eigenvalue problem for the
squared mode frequencies \(\Omega_n^2\).

We use the radial-pulsation equations of Ref.~\cite{Kain:2021rmk},
generalised to the solitonic potential above.  The resulting
eigenvalue problem is solved by a regular-centre shooting scheme with
a finite-radius boundary determinant.  Mode identities are checked
from the reconstructed eigenfunctions and their behaviour under
changes of the outer boundary.  The explicit perturbation equations,
boundary conditions, numerical procedure, and validation tests are
given in the Appendix.

Modes with \(\Omega_n^2>0\) are oscillatory, whereas
\(\Omega_n^2<0\) corresponds to an exponentially growing radial
perturbation.  We therefore define the radial instability index
\begin{equation}
    \nu=\#\{n:\Omega_n^2<0\},
\end{equation}
as the number of unstable radial modes.  This classification follows
directly from the linear spectrum and does not rely on nonlinear time
evolution.  The lowest mode, \(n=0\), is the fundamental radial mode.
In the endpoint region studied below, the first overtone remains
positive, \(\Omega_1^2>0\).  The local stability changes are therefore
set by the zero crossings of \(\Omega_0^2\), with \(\nu\) changing
between \(0\) and \(1\).

\section{Disappearance of the radially stable region}    

\begin{figure}[t]
    \centering
    \includegraphics[width=0.85\columnwidth]{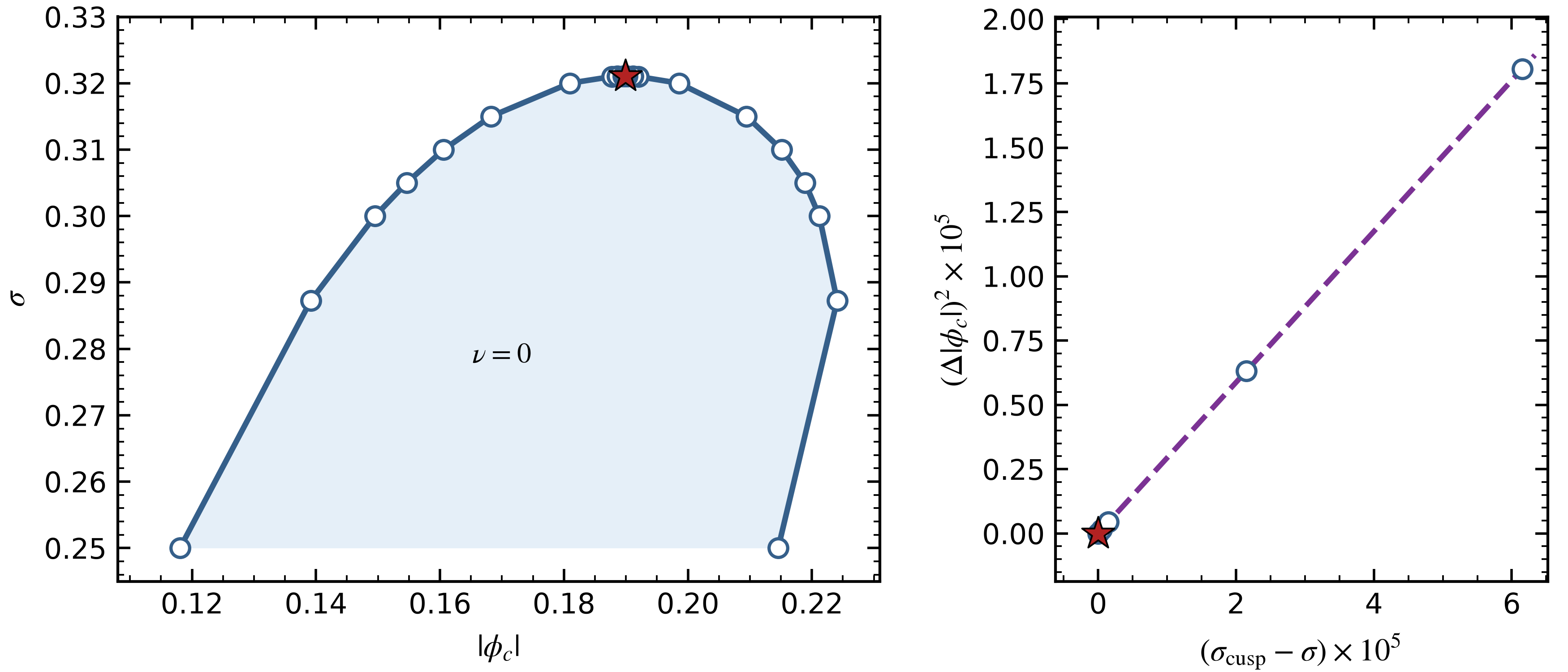}
    \caption{
    Radial stability and approach to the endpoint.
    The left panel shows the radially stable region
    (\(\nu=0\)) in the displayed portion of the
    \((|\phi_c|,\sigma)\) plane, bounded in \(|\phi_c|\) by two
    fundamental zero-mode curves that coalesce at the endpoint.
    The star marks their common endpoint.
    The right panel shows the squared boundary separation
    \((\Delta|\phi_c|)^2\) against
    \(\sigma_{\mathrm{cusp}}-\sigma\); the dashed line is the linear
    near-endpoint fit, and the star marks the endpoint limit.
}
    \label{fig:main1}
\end{figure}

For each fixed \(\sigma\), we bracket and refine the zero crossings of
the fundamental eigenvalue \(\Omega_0^2\) as \(|\phi_c|\) is varied,
and then follow these crossings across the two-parameter equilibrium
family.  The resulting stability map is shown in the left panel of
Fig.~\ref{fig:main1}.  For
\(\sigma<\sigma_{\mathrm{cusp}}\), two curves,
\(|\phi_c|_-(\sigma)\) and \(|\phi_c|_+(\sigma)\), on which the
fundamental radial mode has zero frequency enclose a finite radially
stable region with \(\nu=0\).  These are spectral stability
boundaries, rather than boundaries inferred from extrema of an
equilibrium quantity.

On either side of this interval, within the branch structure shown,
\(\nu=1\), so that along a fixed-\(\sigma\) sequence
\begin{equation}
    \nu:1\longrightarrow0\longrightarrow1 .
\end{equation}
As \(\sigma\) is increased, the two boundaries approach one another
and meet at
\begin{equation}
    \sigma_{\mathrm{cusp}}\simeq0.3210615108,
    \qquad
    |\phi_c|_{\mathrm{cusp}}\simeq0.189896483 .
\end{equation}
For \(\sigma>\sigma_{\mathrm{cusp}}\), no nearby \(\nu=0\) interval
remains.

The approach to the endpoint is quantified in the right panel of
Fig.~\ref{fig:main1}.  Defining
\(\Delta|\phi_c|\equiv|\phi_c|_+-|\phi_c|_-\), we find near the
endpoint
\begin{equation}
    \bigl(\Delta|\phi_c|\bigr)^2
    \simeq
    0.29367\,
    (\sigma_{\mathrm{cusp}}-\sigma),
\end{equation}
and hence
\(\Delta|\phi_c|
\propto(\sigma_{\mathrm{cusp}}-\sigma)^{1/2}\).
The two radial-stability boundaries therefore meet with a square-root
scaling consistent with a cusp-type coalescence.  This scaling
characterises the geometric closing of the stable island; by itself,
however, it does not establish a higher-order critical endpoint.
We therefore next examine the radial spectrum and the constrained variational response

\section{Spectral and variational structure}

\begin{figure}[t]
    \centering
    \includegraphics[width=0.75\columnwidth]{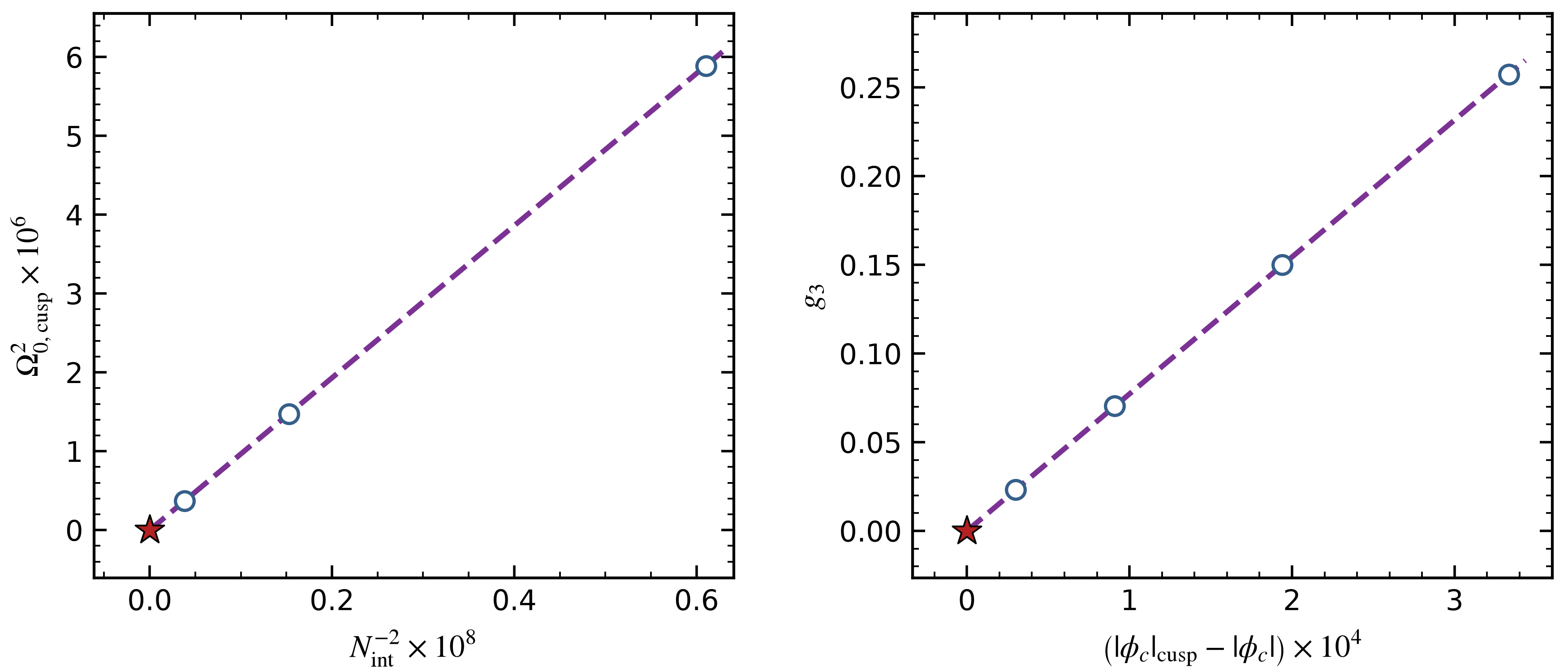}
    \caption{
    Spectral and variational structure of the endpoint.
    The left panel shows the finite-resolution fundamental eigenvalue
    \(\Omega_{0,\mathrm{cusp}}^2\) against \(N_{\rm int}^{-2}\);
    the dashed line gives the continuum extrapolation, with
    \(N_{\rm int}^{-2}\to0\) corresponding to the continuum limit.
    The right panel shows the cubic reduced coefficient \(g_3\) along
    the lower-\(|\phi_c|\) zero-mode boundary as the endpoint is
    approached; the dashed line shows the near-endpoint fit, and the
    star marks \(g_3=0\) at the endpoint.
    }
    \label{fig:main2}
\end{figure}

We first examine the continuum behaviour of the fundamental radial
eigenvalue at the coalescence point identified in
Fig.~\ref{fig:main1}.  The endpoint is located independently at each
radial resolution.  The left panel of Fig.~\ref{fig:main2} shows the
corresponding finite-grid value of the fundamental eigenvalue.
Writing \(N_{\rm int}\) for the radial resolution, we find
\begin{equation}
    \Omega_{0,\mathrm{cusp}}^2
    \propto
    N_{\rm int}^{-2},
\end{equation}
and hence
\begin{equation}
    \Omega_{0,\mathrm{cusp}}^2
    \xrightarrow[N_{\rm int}\to\infty]{}
    0 .
\end{equation}
The small positive value at finite resolution is therefore consistent
with discretisation error rather than a finite spectral gap.

We next determine the order of the degeneracy. 
Because the Noether charge \(Q\) is conserved, the relevant
variational problem at fixed \(\sigma\) is the extremisation of the
ADM mass \(M\) at fixed \(Q\).
We therefore use
the constrained stationary functional
\begin{equation}
    \mathcal F[u;\omega]
    =
    M[u]-\omega Q[u],
\end{equation}
where \(u\) denotes a field configuration and \(\omega\) is the
Lagrange multiplier enforcing fixed \(Q\), equal to the stationary
scalar-field frequency introduced above.

Since the first radial overtone remains positive in the endpoint
region, only the fundamental radial mode becomes soft.  The constrained
functional may therefore be reduced locally to this single critical
direction,
\begin{equation}
    F_{\rm red}(a)
    =
    F_{\mathrm{cusp}}
    +\frac{\lambda}{2}a^2
    +\frac{g_3}{3!}a^3
    +\frac{g_4}{4!}a^4
    +\cdots ,
\end{equation}
where \(F_{\mathrm{cusp}}\) is the value of the reduced functional at
the endpoint, \(\lambda\) is the quadratic stiffness, and \(g_3\) and
\(g_4\) are the cubic and quartic reduced coefficients.  We choose the
normalisation of the soft coordinate \(a\) so that its restriction to
the equilibrium family near the endpoint is
\begin{equation}
    a=|\phi_c|-|\phi_c|_{\mathrm{cusp}} .
\end{equation}

For the equilibrium sequence at fixed \(\sigma\), we define
\begin{equation}
    Q'
    \equiv
    \left.
    \frac{\partial Q}{\partial|\phi_c|}
    \right|_{\sigma},
    \qquad
    \omega'
    \equiv
    \left.
    \frac{\partial\omega}{\partial|\phi_c|}
    \right|_{\sigma}.
\end{equation}
Higher derivatives, including \(Q''\) and \(Q'''\), are defined
analogously.  At the radial zero-mode turning points considered here,
the constrained quadratic stiffness vanishes, \(\lambda=0\), while the
equilibrium sequence satisfies \(Q'=0\).  Using the equilibrium first
law
\begin{equation}
    dM=\omega\,dQ
\end{equation}
along a fixed-\(\sigma\) sequence, the cubic coefficient is
\begin{equation}
    g_3=-\omega'Q'' .
\end{equation}
At an ordinary turning point, \(\omega'\neq0\) and \(Q''\neq0\), so
\(g_3\neq0\).  The quadratic stiffness has vanished, but the cubic term
remains as the leading nonlinear contribution along the soft direction.

The endpoint is more degenerate.  There the two zero-mode turning
points coalesce and
\begin{equation}
    Q'=Q''=0 .
\end{equation}
The additional condition \(Q''=0\) removes the ordinary
turning-point curvature and hence
\begin{equation}
    g_3=0 .
\end{equation}
The next reduced coefficient is
\begin{equation}
    g_4=-\omega'Q''' .
\end{equation}
Since \(Q'''\neq0\) and \(\omega'\neq0\) at the endpoint, \(g_4\)
remains finite and non-zero.  The degeneracy therefore terminates at
quartic order.

The right panel of Fig.~\ref{fig:main2} shows the numerical approach
to this higher-order point.  Along the lower-\(|\phi_c|\) zero-mode
boundary, \(g_3\) approaches zero as the endpoint is approached, while
the quartic coefficient remains finite,
\begin{equation}
    g_4\simeq-775.77 .
\end{equation}
A coalescing-turning-point estimate and an independent fixed-\(Q\)
off-shell reduction give consistent values; the numerical details are
given in the Appendix.

Combining the continuum extrapolation with the constrained reduction,
the endpoint satisfies
\begin{equation}
    \Omega_0^2=0 ,
\end{equation}
together with
\begin{equation}
    \lambda=0,
    \qquad
    g_3=0,
    \qquad
    g_4\neq0 .
\end{equation}
Here \(\Omega_0^2=0\) is the spectral statement that the fundamental
radial mode becomes soft, whereas \(\lambda=0\) states that the
quadratic stiffness of the reduced functional vanishes.  They refer to
the same soft direction, but no numerical identification of
\(\lambda\) with \(\Omega_0^2\) is assumed.  The endpoint is therefore
not distinguished merely by the presence of a zero mode: the leading
cubic response of an ordinary turning point vanishes there as well,
leaving the quartic term as the first non-zero contribution.  This is
the higher-order degeneracy of the constrained radial dynamics.

\section{Local equilibrium structure} 

We now determine how the nearby stationary solutions are organised
around this degeneracy.  Using the same coordinate \(a\), we write
\begin{equation}
    s=\sigma-\sigma_{\mathrm{cusp}} .
\end{equation}
For a nearby fixed charge \(Q_0\), define the charge offset after
subtracting the linear drift of the endpoint,
\begin{equation}
    \Delta_Q
    \equiv
    Q_0-Q_{\mathrm{cusp}}-Q_\sigma s ,
\end{equation}
where
\(Q_\sigma=(\partial Q/\partial\sigma)_{\mathrm{cusp}}\).
Within the local cusp scaling regime,
\(a=O(|s|^{1/2})\) and
\(\Delta_Q=O(|s|^{3/2})\).
The terms \(sa\) and \(a^3\) therefore enter at the same leading order,
whereas \(s^2\), \(a^2s\), and \(a^4\) are higher order.  Since the
endpoint also satisfies \(Q'=Q''=0\), the fixed-charge condition
\(Q=Q_0\) becomes
\begin{equation}
    0
    =
    -\Delta_Q
    +
    Q_{|\phi_c|\sigma}\,s\,a
    +
    \frac{Q'''}{6}a^3
    +\cdots ,
\end{equation}
where
\begin{equation}
    Q_{|\phi_c|\sigma}
    \equiv
    \left.
    \frac{\partial^2 Q}
    {\partial|\phi_c|\,\partial\sigma}
    \right|_{\mathrm{cusp}} .
\end{equation}
Since \(Q'''\neq0\) and \(Q_{|\phi_c|\sigma}\neq0\), this can be
written as
\begin{equation}
    a^3+p\,a+q=0 ,
\end{equation}
with
\begin{equation}
    p
    =
    \frac{6Q_{|\phi_c|\sigma}}{Q'''}\,s,
    \qquad
    q
    =
    -\frac{6\Delta_Q}{Q'''} .
\end{equation}
For the coefficients found here, \(p\) has the same sign as \(s\).
Thus, for \(\sigma<\sigma_{\mathrm{cusp}}\), a finite range of nearby
charges admits three nearby stationary solutions.  The cubic equation
determines only their local multiplicity; their radial stability is
fixed independently by the pulsation spectrum, which gives the
\(\nu:1\to0\to1\) structure found above.  This three-solution region
shrinks to the endpoint as
\(\sigma\to\sigma_{\mathrm{cusp}}\), whereas for
\(\sigma>\sigma_{\mathrm{cusp}}\) only one nearby real solution
remains.  The equilibrium family therefore has the local
two-parameter structure of a non-degenerate cusp.  Together with the
radial zero mode and the quartic constrained reduction, this identifies
the endpoint as a cusp-type higher-order critical endpoint.

\section{Critical softening}

The approach to the endpoint is also visible directly in the
fundamental radial spectrum.  For each
\(\sigma<\sigma_{\mathrm{cusp}}\), we evaluate the lowest radial mode
at the midpoint
\[
    |\phi_c|_{\mathrm{mid}}
    =
    \frac{|\phi_c|_-+|\phi_c|_+}{2},
\]
which follows the centre of the shrinking stable interval up to
higher-order corrections.  This avoids the zero-mode boundaries
themselves, where \(\Omega_0^2=0\) by construction, and instead tracks
the characteristic spectral gap inside the stable region.

At each radial resolution \(N_{\rm int}\), we subtract the
independently measured finite-grid cusp residual,
\begin{equation}
    \Delta\Omega_0^2(N_{\rm int},\sigma)
    =
    \Omega_0^2(N_{\rm int},\sigma)
    -
    \Omega_{0,\mathrm{cusp}}^2(N_{\rm int}) .
\end{equation}
The three resolutions give consistent slopes.  Using the converged
value, we obtain
\begin{equation}
    \Delta\Omega_0^2
    \simeq
    0.49746\,
    (\sigma_{\mathrm{cusp}}-\sigma).
\end{equation}
Defining the corresponding soft-mode frequency by
\(\Omega_{\mathrm{soft}}\equiv\sqrt{\Delta\Omega_0^2}\), this gives
the square-root critical law
\begin{equation}
    \Omega_{\mathrm{soft}}
    \simeq
    0.70531\,
    (\sigma_{\mathrm{cusp}}-\sigma)^{1/2}.
\end{equation}
Here \(\Omega_{\mathrm{soft}}\) is the residual-corrected soft-mode
frequency; it coincides with the fundamental radial frequency in the
continuum limit.  The corresponding linear soft-mode timescale,
\(\tau_{\mathrm{soft}}\equiv\Omega_{\mathrm{soft}}^{-1}\),
therefore diverges as
\((\sigma_{\mathrm{cusp}}-\sigma)^{-1/2}\).
The disappearance of the stable region is thus accompanied by
critical softening of the fundamental radial mode, rather than only
by the coalescence of stationary turning points.

\section{Robustness under potential deformation}    

We next test whether the endpoint depends sensitively on the precise
form of the self-interaction.  We deform only the sextic term,
\begin{equation}
    V_\epsilon(\phi)
    =
    |\phi|^2
    -\frac{4|\phi|^4}{\sigma^2}
    +\frac{4(1+\epsilon)|\phi|^6}{\sigma^4},
    \qquad
    -0.10\leq\epsilon\leq0.10 .
\end{equation}
Across this interval the endpoint persists and moves smoothly in the
\((|\phi_c|,\sigma)\) plane.  The quantities \(Q'''\) and
\(\omega'\) remain non-zero, while \(g_4\) stays finite and does not
change sign.  The finite-resolution radial eigenvalue at the deformed
endpoint remains small throughout the tested range, and second-order
convergence towards zero is recovered for the representative
deformations \(\epsilon=\pm0.05\), as well as for the undeformed case.
The cusp-type critical structure is therefore numerically robust under
the tested non-trivial deformation of the scalar potential.  Further
details are given in the Appendix.

\section{Conclusion}    

We have shown that a finite radially stable region in the
two-parameter family of solitonic boson stars terminates at a
cusp-type endpoint.  The two boundaries defined by
\(\Omega_0^2=0\) coalesce there.  In the constrained reduction, the
quadratic stiffness and cubic coefficient vanish,
\(\lambda=g_3=0\), while the quartic coefficient remains finite and
non-zero.  This distinguishes the endpoint from an ordinary radial
zero-mode turning point.

The local fixed-charge equilibrium condition reduces to a
non-degenerate cubic cusp structure, consistent with the independently
resolved \(\nu:1\to0\to1\) radial-stability pattern below the endpoint.
Approaching the cusp from below, the fundamental mode softens as
\(\Omega_{\mathrm{soft}}
\propto(\sigma_{\mathrm{cusp}}-\sigma)^{1/2}\), so that its associated
linear timescale diverges.  The endpoint also persists under the
tested deformation of the sextic self-interaction.

The disappearance of the stable island is accompanied by a
progressive softening of the fundamental radial mode. Since both
the island width \(\Delta|\phi_c|\) and the soft-mode frequency
\(\Omega_{\mathrm{soft}}\) scale as
\((\sigma_{\mathrm{cusp}}-\sigma)^{1/2}\), we find
\(\Omega_{\mathrm{soft}}\propto\Delta|\phi_c|\) near the endpoint.
Together with the vanishing cubic coefficient and the finite
quartic term in the fixed-charge reduction, this relates the
closing of the stable region to the loss of radial restoring
force at the endpoint.

\begin{acknowledgments}
The author acknowledges HIAS for access to the ``Quantum Universe
Physical Simulation Platform''. This work used the DiRAC Memory
Intensive service (COSMA8) at Durham University, managed by the
Institute for Computational Cosmology on behalf of the STFC DiRAC HPC
Facility (www.dirac.ac.uk). The DiRAC service at Durham was funded by
BEIS, UKRI and STFC capital funding, Durham University and STFC
operations grants. DiRAC is part of the UKRI Digital Research
Infrastructure. This work was supported by the National Natural Science
Foundation of China under Grant No.~12505066.
\end{acknowledgments}

\clearpage
\appendix

\section{Radial pulsation problem and numerical procedure}

\subsection{Radial perturbation equations}

The stationary backgrounds are obtained with the shooting construction
described in Ref.~\cite{Helfer:2021brt}.  The baseline equilibrium
atlas uses a uniform radial grid with
\begin{equation}
    R_{\max}=200,
    \qquad
    N_{\rm int}=12801 ,
    \label{eq:supp_baseline_resolution}
\end{equation}
where \(N_{\rm int}\) is the number of radial grid points.  The endpoint
calculations are repeated at \(N_{\rm int}=25601\) and \(51201\).

For the radial calculation we follow the perturbation variables of
Ref.~\cite{Kain:2021rmk}.  The stationary fields are mapped to those
conventions according to
\begin{equation}
    \varphi_0
    =
    \frac{|\phi|}{\sqrt{2}},
    \qquad
    \nu_0=2\Phi,
    \qquad
    \lambda_0=2\ln X ,
    \label{eq:supp_kain_map}
\end{equation}
so that
\begin{equation}
    ds^2
    =
    -e^{\nu_0}dt^2
    +e^{\lambda_0}dr^2
    +r^2d\Omega^2 .
\end{equation}
Writing \(s=|\varphi|^2\), the potential in these conventions is
\begin{equation}
    U(s)
    =
    \left.
    \frac{1}{2}V
    \right|_{|\phi|^2=2s}
    =
    s
    \left(
        1-\frac{4s}{\sigma^2}
    \right)^2 .
    \label{eq:supp_Umap}
\end{equation}
Its derivatives are
\begin{align}
    U_s
    &=
    \left(
        1-\frac{4s}{\sigma^2}
    \right)
    \left(
        1-\frac{12s}{\sigma^2}
    \right),
    \label{eq:supp_Us}
    \\
    U_{ss}
    &=
    -\frac{16}{\sigma^2}
    +\frac{96s}{\sigma^4}.
    \label{eq:supp_Uss}
\end{align}

In these conventions the perturbed field is written as
\begin{equation}
    \varphi(t,r)
    =
    \left[
        \varphi_1(t,r)
        +i\varphi_2(t,r)
    \right]e^{-i\omega t},
\end{equation}
with
\begin{align}
    \varphi_1(t,r)
    &=
    \varphi_0(r)
    \left[
        1+\delta\varphi_1(t,r)
    \right],
    \\
    \varphi_2(t,r)
    &=
    \varphi_0(r)\delta\varphi_2(t,r),
    \\
    \nu(t,r)
    &=
    \nu_0(r)+\delta\nu(t,r),
    \\
    \lambda(t,r)
    &=
    \lambda_0(r)+\delta\lambda(t,r).
    \label{eq:supp_perturbations}
\end{align}
The opposite sign of the stationary phase relative to the convention
used in the main text is related by complex conjugation and does not
change the spectrum.  Following Ref.~\cite{Kain:2021rmk}, we introduce
\begin{equation}
    \dot{\xi}
    =
    \omega\,\delta\varphi_2,
    \qquad
    \zeta=\xi',
    \label{eq:supp_zeta_def}
\end{equation}
and assume harmonic time dependence with radial eigenfrequency
\(\Omega\).  Only \(\Omega^2\) enters the radial equations.

In this section primes denote radial derivatives.  For compactness,
define
\begin{equation}
    f\equiv\delta\varphi_1,
    \qquad
    h\equiv\frac{\varphi_0'}{\varphi_0},
    \qquad
    E\equiv e^{\lambda_0-\nu_0}.
    \label{eq:supp_radial_defs}
\end{equation}
Carrying the linearisation of Ref.~\cite{Kain:2021rmk} through for a
general \(U(|\varphi|^2)\) gives
\begin{align}
    f''
    ={}&
    -\Omega^2Ef
    -
    \left(
        2h+\frac{2}{r}
        +\frac{\nu_0'-\lambda_0'}{2}
    \right)f'
    +2\mathcal B\,\zeta
    +2\zeta'
    -
    16\pi r\varphi_0
    \left(
        \varphi_0'f-\varphi_0\zeta
    \right)
    \mathcal C
    \nonumber\\
    &+
    \biggl[
        4\omega^2E
        +
        16\pi r e^{\lambda_0}
        \varphi_0\varphi_0'
        \left(
            \omega^2e^{-\nu_0}+U_s
        \right)
        +
        2e^{\lambda_0}\varphi_0^2U_{ss}
    \biggr]f ,
    \label{eq:supp_puls_f}
\end{align}
and
\begin{align}
    \zeta''
    ={}&
    -\Omega^2E\zeta
    -2\omega^2Ef'
    -
    \mathcal B
    \left[
        (\nu_0'-\lambda_0')\zeta+\zeta'
    \right]
    -\mathcal D\zeta
    -(\nu_0'-\lambda_0')\zeta'
    \nonumber\\
    &-
    8\pi\omega^2E
    \biggl[
        \left(
            \varphi_0^2
            +2r\varphi_0\varphi_0'
        \right)\zeta
        +r\varphi_0^2\zeta'
        -r\varphi_0
        \left(
            \nu_0'-\lambda_0'
            +\frac{2}{r}
        \right)
        \left(
            \varphi_0'f-\varphi_0\zeta
        \right)
        +
        2r e^{\lambda_0}
        \varphi_0^2U_s f
    \biggr],
    \label{eq:supp_puls_zeta}
\end{align}
where
\begin{align}
    \mathcal B
    &=
    \frac{2}{r}
    +2h
    -\lambda_0'
    +8\pi r(\varphi_0')^2,
    \label{eq:supp_B}
    \\
    \mathcal C
    &=
    \frac{h}{2}
    \left(
        \nu_0'-\lambda_0'
        +\frac{2}{r}
    \right)
    +
    e^{\lambda_0}
    \left(
        \omega^2e^{-\nu_0}-U_s
    \right),
    \label{eq:supp_C}
    \\
    \mathcal D
    &=
    \frac{2\varphi_0''}{\varphi_0}
    -2h^2
    -\frac{2}{r^2}
    -\lambda_0''
    +
    8\pi
    \left[
        (\varphi_0')^2
        +2r\varphi_0'\varphi_0''
    \right].
    \label{eq:supp_D}
\end{align}
The second derivative \(\varphi_0''\) is evaluated from the stationary
Klein--Gordon equation.  We evaluate \(\lambda_0'\) from the background
metric equations and obtain \(\lambda_0''\) by differentiating
\(\lambda_0'\) on the radial grid.

The metric and enclosed-charge perturbations needed below are
\begin{align}
    \delta\lambda
    &=
    16\pi r\varphi_0
    \left(
        \varphi_0'f-\varphi_0\zeta
    \right),
    \label{eq:supp_dlambda}
    \\
    \delta q
    &=
    -\frac{8\pi}{\omega}
    r^2\varphi_0^2
    e^{(\nu_0-\lambda_0)/2}
    \zeta .
    \label{eq:supp_dq}
\end{align}

\subsection{Shooting determinant, mode identification, and validation}

Regularity at the centre leaves two independent amplitudes.  We write
the leading terms as
\begin{align}
    f(r)
    &=
    f_0+c_2r^2+O(r^4),
    \\
    \zeta(r)
    &=
    \zeta_1r+O(r^3),
    \label{eq:supp_center_expansion}
\end{align}
where
\begin{equation}
    c_2
    =
    \zeta_1
    +
    \frac{f_0}{6}
    \left[
        \frac{4\omega^2-\Omega^2}{\alpha_c^2}
        +
        2\varphi_{0c}^2U_{ss,c}
    \right],
    \label{eq:supp_c2}
\end{equation}
with
\begin{equation}
    \alpha_c=e^{\Phi(0)},
    \qquad
    \varphi_{0c}
    =
    \frac{|\phi_c|}{\sqrt{2}},
    \qquad
    U_{ss,c}
    \equiv
    U_{ss}(\varphi_{0c}^{\,2}).
\end{equation}
For each trial \(\Omega^2\), we integrate two regular basis solutions
with \((f_0,\zeta_1)=(1,0)\) and \((0,1)\).

The physical outer conditions are
\begin{equation}
    \varphi_0 f\rightarrow0,
    \qquad
    \delta\lambda\rightarrow0,
    \qquad
    \delta q\rightarrow0
    \qquad
    (r\rightarrow\infty).
    \label{eq:supp_pert_outer}
\end{equation}
The condition \(\delta q(\infty)=0\) enforces vanishing first-order
variation of the total Noether charge.  At a finite outer radius
\(r_{\rm out}\) in the decaying scalar region we impose
\begin{equation}
    f(r_{\rm out})=0,
    \qquad
    \zeta(r_{\rm out})=0 .
    \label{eq:supp_finite_outer}
\end{equation}
Through Eqs.~\eqref{eq:supp_dlambda} and \eqref{eq:supp_dq}, these
conditions also set \(\delta\lambda\) and \(\delta q\) to zero at
\(r_{\rm out}\).

When the stationary scalar matching radius \(r_{\rm match}\), at which
the numerical scalar profile is joined to its exterior continuation,
is available, the outer boundaries are chosen before it,
\begin{equation}
    r_{\rm out}
    =
    (0.68,\ 0.78,\ 0.87,\ 0.93)\,r_{\rm match},
    \label{eq:supp_rout}
\end{equation}
so that the pulsation equations use the directly integrated stationary
scalar profile.  If \(r_{\rm match}\) is unavailable, we use the radius
\(R_{99}\) enclosing \(99\%\) of the ADM mass and take
\begin{equation}
    r_{\rm out}
    =
    (1.2,\ 1.4,\ 1.6,\ 1.8)\,R_{99}.
\end{equation}
Fixed fallback radii \(18,22,26,30\) are used only when neither scale
is available.  In all cases the remaining outer-boundary dependence
is checked explicitly.

Let \(Y^{(1)}\) and \(Y^{(2)}\) denote the two regular basis solutions.
We form
\begin{equation}
    \mathsf B(\Omega^2;r_{\rm out})
    =
    \begin{pmatrix}
        f^{(1)}(r_{\rm out})
        &
        f^{(2)}(r_{\rm out})
        \\
        \zeta^{(1)}(r_{\rm out})
        &
        \zeta^{(2)}(r_{\rm out})
    \end{pmatrix}.
    \label{eq:supp_boundary_matrix}
\end{equation}
An eigenvalue is a zero of the shooting determinant,
\begin{equation}
    D(\Omega^2;r_{\rm out})
    \equiv
    \det\mathsf B(\Omega^2;r_{\rm out})
    =0 .
    \label{eq:supp_det}
\end{equation}
The determinant is scanned in \(\Omega^2\), sign-changing roots are
bracketed, and each root is refined with Brent's method.  If the
lowest root in the initial interval is not the fundamental mode, the
negative search range is extended until a nodeless root is found.
The required linear combination of the two regular basis solutions is
then obtained from the null vector of \(\mathsf B\) by singular-value
decomposition.

Mode identity is established from eigenvalue ordering and continuity
along neighbouring equilibrium solutions, together with stability
under changes of \(r_{\rm out}\).  The node count of
\(\delta\lambda\) provides an additional diagnostic: the fundamental
mode is nodeless and the first overtone has one radial node.  Nodes
confined to the numerically negligible far tail are discarded.  The
node count is also used to detect cases in which the initial negative
search interval has missed the fundamental root.

The pulsation equations are integrated with fourth-order Runge--Kutta
on the stationary radial grid, with background quantities linearly
interpolated at the intermediate Runge--Kutta stages.  A mode is
retained only when its eigenvalue and identity are stable as
\(r_{\rm out}\) is increased.

As an implementation check, the solver was specialised to the
mini-boson-star model studied in Ref.~\cite{Kain:2021rmk}.  After
mapping conventions, the fundamental eigenvalue agrees with the
published result at the \(10^{-4}\) relative level, with the difference
decreasing as the outer boundary is moved outwards.  In the endpoint
region, both the fundamental and first-overtone branches are resolved.
The first overtone remains positive, so the local changes in the radial
instability index are due solely to zero crossings of the fundamental
mode.

\section{Endpoint location and continuum convergence}

\subsection{Tracing the zero-mode boundaries}

At fixed \(\sigma\), the two boundaries of the stable interval are
located directly from the fundamental radial spectrum by solving
\begin{equation}
    \Omega_0^2(|\phi_c|,\sigma)=0 .
    \label{eq:supp_zero_boundary}
\end{equation}
Starting where the two zero crossings are well separated, we bracket
and refine each root in \(|\phi_c|\) and continue both towards larger
\(\sigma\).  We denote the resulting branches by
\begin{equation}
    |\phi_c|_-(\sigma),
    \qquad
    |\phi_c|_+(\sigma),
    \qquad
    |\phi_c|_-<|\phi_c|_+ .
    \label{eq:supp_two_boundaries}
\end{equation}
In the branch segment relevant to the endpoint,
\(\Omega_0^2>0\) between these roots and
\(\Omega_0^2<0\) on either side.  The boundaries are therefore defined
from the radial spectrum rather than inferred from extrema of an
equilibrium quantity.

Their separation,
\begin{equation}
    \Delta|\phi_c|
    \equiv
    |\phi_c|_+-|\phi_c|_- ,
    \label{eq:supp_boundary_separation}
\end{equation}
shrinks continuously on approaching the endpoint.  In the
near-endpoint regime,
\begin{equation}
    \left(\Delta|\phi_c|\right)^2
    \simeq
    0.29367\,
    (\sigma_{\rm cusp}-\sigma),
    \label{eq:supp_separation_scaling}
\end{equation}
so that
\(\Delta|\phi_c|\propto(\sigma_{\rm cusp}-\sigma)^{1/2}\).
The zero of the linear fit to
\(\left(\Delta|\phi_c|\right)^2\) provides one estimate of the
endpoint.

\subsection{Endpoint refinement and local non-degeneracy}

We refine the endpoint using two complementary quantities that vanish
at the coalescence.  Besides the boundary separation above, we use the
cubic coefficient \(g_3\) of the constrained reduction, derived in the
next section.  Along the lower-\(|\phi_c|\) zero-mode branch used for
the fit, the near-endpoint behaviour is
\begin{equation}
    g_3^2
    \propto
    \sigma_{\rm cusp}-\sigma .
    \label{eq:supp_g3_endpoint_scaling}
\end{equation}
Thus \(g_3^2\) provides a second extrapolation to the same endpoint.

At \(N_{\rm int}=25601\), the two estimates are
\begin{align}
    \sigma_{\rm cusp}^{(g_3^2)}
    &=
    0.321061510914463,
    \\
    \sigma_{\rm cusp}^{(\Delta|\phi_c|)^2}
    &=
    0.321061510893534 ,
\end{align}
and differ by \(2.09\times10^{-11}\).  At
\(N_{\rm int}=51201\),
\begin{align}
    \sigma_{\rm cusp}^{(g_3^2)}
    &=
    0.321061510843676,
    \\
    \sigma_{\rm cusp}^{(\Delta|\phi_c|)^2}
    &=
    0.321061510822502 ,
\end{align}
with a difference of \(2.12\times10^{-11}\).  This close agreement is
a stringent internal check on the endpoint refinement, but is not
interpreted as a continuum error estimate.

The working endpoint coordinates at each resolution are listed in
Table~\ref{tab:cusp_convergence}.  We quote
\begin{equation}
    \sigma_{\rm cusp}\simeq0.3210615108,
    \qquad
    |\phi_c|_{\rm cusp}\simeq0.189896483 ,
    \label{eq:supp_cusp_coordinates}
\end{equation}
without assigning significance to the final digits beyond the observed
resolution dependence.

\begin{table}[t]
    \centering
    \caption{
    Resolution dependence of the refined endpoint and of the
    fundamental radial eigenvalue evaluated there.  The residual
    positive value of \(\Omega_{0,\rm cusp}^{2}\) decreases by
    approximately a factor of four under each doubling of the radial
    resolution.
    }
    \label{tab:cusp_convergence}
    \begin{ruledtabular}
    \begin{tabular}{cccc}
        \(N_{\rm int}\)
        & \(\sigma_{{\rm cusp},N}\)
        & \(|\phi_c|_{{\rm cusp},N}\)
        & \(\Omega_{0,\rm cusp}^{2}\)
        \\
        \hline
        12801
        & 0.321061512300000
        & 0.189896480200000
        & \(5.891183594817\times10^{-6}\)
        \\
        25601
        & 0.321061510903998
        & 0.189896483560000
        & \(1.473369209294\times10^{-6}\)
        \\
        51201
        & 0.321061510833089
        & 0.189896482950000
        & \(3.683769285134\times10^{-7}\)
    \end{tabular}
    \end{ruledtabular}
\end{table}

The endpoint is refined independently at each resolution before the
radial eigenvalue is evaluated.  The successive ratios are
\begin{equation}
    \frac{\Omega_{0,\rm cusp}^2(12801)}
         {\Omega_{0,\rm cusp}^2(25601)}
    \simeq3.99844,
    \qquad
    \frac{\Omega_{0,\rm cusp}^2(25601)}
         {\Omega_{0,\rm cusp}^2(51201)}
    \simeq3.99962 ,
    \label{eq:supp_cusp_ratio}
\end{equation}
corresponding to convergence orders
\begin{equation}
    p_{\rm conv}\simeq1.99944,
    \qquad
    p_{\rm conv}\simeq1.99986 .
    \label{eq:supp_cusp_order}
\end{equation}
Thus
\begin{equation}
    \Omega_{0,\rm cusp}^2
    \propto
    N_{\rm int}^{-2},
    \label{eq:supp_cusp_radial_convergence}
\end{equation}
and the fundamental radial eigenvalue is consistent with zero in the
continuum limit.  The residual positive value at finite resolution is
therefore treated as a discretisation effect rather than a physical
spectral gap.

From this point on, primes on the equilibrium quantities \(Q\) and
\(\omega\) denote differentiation with respect to \(|\phi_c|\) at
fixed \(\sigma\).  The local fit at the endpoint is consistent with
\begin{equation}
    Q'=0,
    \qquad
    Q''=0 .
    \label{eq:supp_cusp_degeneracy}
\end{equation}
A representative fit at \(N_{\rm int}=51201\) gives
\begin{equation}
    Q'''
    \simeq
    -535.785905,
    \qquad
    Q_{|\phi_c|\sigma}
    \simeq
    -19.668171 ,
    \label{eq:supp_cusp_nondegeneracy}
\end{equation}
where
\begin{equation}
    Q_{|\phi_c|\sigma}
    \equiv
    \left.
    \frac{\partial^2 Q}
    {\partial|\phi_c|\,\partial\sigma}
    \right|_{\rm cusp}.
\end{equation}
The non-zero values of \(Q'''\) and \(Q_{|\phi_c|\sigma}\) satisfy the
local non-degeneracy conditions for the equilibrium cusp.  The
frequency derivative is also finite,
\begin{equation}
    \omega'
    \simeq
    -1.44790858 ,
    \label{eq:supp_omega_derivative}
\end{equation}
and enters the constrained reduction below.

\section{Constrained reduction and quartic closure}

\subsection{Constrained tangent and soft direction}

At fixed \(\sigma\), stationary boson stars are critical points of the
constrained functional
\begin{equation}
    \mathcal F[u;\omega]
    =
    M[u]-\omega Q[u],
    \label{eq:supp_constrained_functional}
\end{equation}
where \(M\) is the ADM mass, \(Q\) the conserved Noether charge,
\(u\) denotes the field configuration, and \(D\) below denotes
functional differentiation with respect to \(u\).  In the present
conventions,
\begin{equation}
    Q
    =
    4\pi\omega
    \int_0^\infty
    r^2 X e^{-\Phi}|\phi|^2\,dr .
    \label{eq:supp_noether_charge}
\end{equation}
The frequency \(\omega\) acts as the Lagrange multiplier enforcing
fixed \(Q\).  Stationarity gives
\begin{equation}
    DM-\omega DQ=0 .
    \label{eq:supp_stationarity}
\end{equation}
Along an equilibrium sequence at fixed \(\sigma\), define
\begin{equation}
    T
    \equiv
    \frac{du}{d|\phi_c|},
    \qquad
    H
    \equiv
    D^2(M-\omega Q).
    \label{eq:supp_HT_def}
\end{equation}
Differentiating Eq.~\eqref{eq:supp_stationarity} along the equilibrium
family gives
\begin{equation}
    H[T,\cdot]
    =
    \omega' DQ[\cdot] .
    \label{eq:supp_HT}
\end{equation}
Since
\begin{equation}
    Q'=DQ[T],
    \label{eq:supp_Qprime}
\end{equation}
the tangent \(T\) lies in the fixed-\(Q\) tangent space when
\(Q'=0\).  For any fixed-\(Q\) tangent perturbation \(v\),
Eq.~\eqref{eq:supp_HT} then gives
\begin{equation}
    H[T,v]=0 .
    \label{eq:supp_constrained_null}
\end{equation}
Thus \(T\) is a null direction of the Hessian restricted to the
fixed-\(Q\) constraint surface.

The pulsation calculation independently finds the fundamental radial
zero mode on the same boundary.  The two descriptions have different
normalisations: the pulsation problem determines \(\Omega_0^2\),
whereas the constrained reduction determines the quadratic stiffness
\(\lambda\).  No numerical identification of these two quantities is
assumed.

The first radial overtone remains positive near the endpoint, so only
the fundamental radial mode softens in the resolved spectrum.  We
therefore use a one-dimensional local reduction, with the coordinate
normalised so that on the equilibrium family
\begin{equation}
    a
    =
    |\phi_c|-|\phi_c|_{\rm cusp}.
    \label{eq:supp_soft_coordinate}
\end{equation}
The reduced functional takes the form
\begin{equation}
    F_{\rm red}(a)
    =
    F_{\rm cusp}
    +\frac{\lambda}{2}a^2
    +\frac{g_3}{3!}a^3
    +\frac{g_4}{4!}a^4
    +O(a^5),
    \label{eq:supp_reduced_functional}
\end{equation}
where \(F_{\rm cusp}\) is its endpoint value and
\(\lambda\), \(g_3\), and \(g_4\) are the quadratic, cubic, and
quartic coefficients.  At the corresponding constrained turning point,
\(\lambda=0\).

\subsection{Cubic and quartic identities}

The coefficients in Eq.~\eqref{eq:supp_reduced_functional} are
obtained from the constrained Lyapunov--Schmidt (LS) reduction of
Eq.~\eqref{eq:supp_stationarity}.  They should not be identified with
derivatives of \(M-\omega_*Q\) taken along the equilibrium family at a
fixed value of \(\omega_*\).

At an ordinary turning point, projection of the twice-differentiated
stationary equation onto the constrained null direction gives
\begin{equation}
    g_3
    =
    -\omega'Q'' .
    \label{eq:supp_g3_identity}
\end{equation}
The sign of \(g_3\) depends on the orientation chosen for the soft
coordinate, while the condition \(g_3=0\) does not.

At the endpoint,
\begin{equation}
    Q'=0,
    \qquad
    Q''=0 ,
    \label{eq:supp_Q_cusp}
\end{equation}
and therefore
\begin{equation}
    g_3=0 .
    \label{eq:supp_g3_zero}
\end{equation}
The condition \(Q''=0\) removes the charge-curvature term that
generates the cubic response at an ordinary turning point.  Carrying
the constrained LS projection one order higher gives
\begin{equation}
    g_4
    =
    -\omega'Q''' .
    \label{eq:supp_g4_identity}
\end{equation}
Since \(\omega'\neq0\) and \(Q'''\neq0\) at the endpoint, the quartic
coefficient remains finite and non-zero.

A separate estimate follows from the two ordinary turning points below
the endpoint.  Using the same orientation
\(a=|\phi_c|-|\phi_c|_{\rm cusp}\) on both branches, their cubic
coefficients are
\begin{equation}
    g_{3,\pm}
    =
    -\omega'Q''
    \big|_{|\phi_c|_\pm}.
    \label{eq:supp_g3_turning_points}
\end{equation}
Near the coalescence,
\begin{equation}
    g_3(|\phi_c|)
    =
    g_4
    \left(
        |\phi_c|-|\phi_c|_{\rm cusp}
    \right)
    +
    O\!\left[
        \left(
            |\phi_c|-|\phi_c|_{\rm cusp}
        \right)^2
    \right],
    \label{eq:supp_g3_near_cusp}
\end{equation}
and hence
\begin{equation}
    g_4
    =
    \lim_{\sigma\to\sigma_{\rm cusp}^{-}}
    \frac{
        g_{3,+}-g_{3,-}
    }{
        |\phi_c|_+-|\phi_c|_-
    } .
    \label{eq:supp_g4_coalescence_estimator}
\end{equation}
This gives a second estimate of \(g_4\) without directly evaluating
\(Q'''\).

\subsection{Numerical derivative extraction}

The derivatives of \(Q(|\phi_c|)\) and \(\omega(|\phi_c|)\) are
extracted from symmetric local polynomial fits about the refined
endpoint at each radial resolution.  The resulting quartic
coefficients are listed in Table~\ref{tab:g4_closure}.

\begin{table}[t]
    \centering
    \caption{
    Quartic coefficient at the undeformed endpoint.  The
    equilibrium-family identity \(g_4=-\omega'Q'''\) is compared with
    the coalescence estimator
    Eq.~\eqref{eq:supp_g4_coalescence_estimator}.
    }
    \label{tab:g4_closure}

    \begin{tabular}{c@{\hspace{2.5em}}c@{\hspace{2.5em}}c}
        \hline\hline
        \(N_{\rm int}\)
        & \(-\omega'Q'''\)
        & \(g_4^{\rm coal}\)
        \\
        \hline
        12801
        & \(-775.7735\)
        & \(-775.7769\)
        \\
        25601
        & \(-775.7702\)
        & \(-775.7712\)
        \\
        51201
        & \(-775.7690\)
        & \(-775.7663\)
        \\
        \hline\hline
    \end{tabular}
\end{table}

The two determinations remain close at all three resolutions and give
\begin{equation}
    g_4\simeq-775.77 .
    \label{eq:supp_g4_value}
\end{equation}
Both its sign and its non-zero character are insensitive to radial
refinement.

As a separate check on the equilibrium data entering the constrained
reduction, we verify the first-law relation
\begin{equation}
    dM=\omega\,dQ .
    \label{eq:supp_first_law}
\end{equation}
At the representative ordinary turning region
\(\sigma=0.28725\), with \(N_{\rm int}=25601\), a symmetric derivative
with step \(5\times10^{-5}\) gives
\begin{equation}
    \max |M'-\omega Q'|
    =
    6.13\times10^{-8},
    \qquad
    \max |\Delta M-\bar{\omega}\Delta Q|
    =
    2.87\times10^{-12}.
    \label{eq:supp_first_law_check}
\end{equation}
Here \(\Delta M\) and \(\Delta Q\) are differences between neighbouring
equilibrium solutions, and \(\bar{\omega}\) is the mean frequency of
the pair.  Doubling the derivative spacing gives comparable residuals.

\subsection{Independent fixed-$Q$ off-shell check}

An independent fixed-\(Q\) calculation determines the quartic
coefficient without using Eq.~\eqref{eq:supp_g4_identity}.  It is
performed at \(N_{\rm int}=25601\) using the endpoint obtained from
the \(g_3^2\) refinement,
\begin{equation}
    \sigma_{\rm cusp}
    =
    0.321061510914,
    \qquad
    |\phi_c|_{\rm cusp}
    =
    0.18989648356 .
    \label{eq:supp_LS_point}
\end{equation}
The small difference between this value of \(\sigma_{\rm cusp}\) and
the working value in Table~\ref{tab:cusp_convergence} reflects the two
endpoint estimators discussed above.

An asterisk denotes evaluation at this refined endpoint, and \(Q_*\)
is its Noether charge.  From symmetric local fits to the equilibrium
profiles we construct
\begin{equation}
    |\phi(r;x)|
    =
    |\phi_*(r)|
    +x\,\chi(r)
    +\frac{x^2}{2}u_2(r),
    \label{eq:supp_offshell_path}
\end{equation}
where
\begin{equation}
    \chi(r)
    =
    \left.
    \frac{\partial|\phi(r)|}
    {\partial|\phi_c|}
    \right|_*,
    \qquad
    u_2(r)
    =
    \left.
    \frac{\partial^2|\phi(r)|}
    {\partial|\phi_c|^2}
    \right|_* .
    \label{eq:supp_chi_u2}
\end{equation}
The coordinate \(x\) is normalised so that
\begin{equation}
    |\phi_c|-|\phi_c|_{\rm cusp}
    =
    x+O(x^2).
    \label{eq:supp_x_normalisation}
\end{equation}
For each \(x\), the scalar profile in
Eq.~\eqref{eq:supp_offshell_path} is prescribed, the Einstein sector
is solved again, and the frequency is adjusted so that
\begin{equation}
    Q(x)=Q_* .
    \label{eq:supp_fixedQ_path}
\end{equation}
The scalar Klein--Gordon equation is deliberately not imposed.  Its
residual \(R_{\rm KG}(r;x)\) therefore measures the departure from
stationarity along this fixed-\(Q\) path.

Projecting the residual along the path tangent gives the reduced
generalised force
\begin{equation}
    \mathcal N(x)
    =
    -4\pi
    \int
    r^2 e^{\Phi(r;x)}
    \left[
        \chi(r)+xu_2(r)
    \right]
    R_{\rm KG}(r;x)\,dr .
    \label{eq:supp_generalised_force}
\end{equation}
Its local expansion is
\begin{equation}
    \mathcal N(x)
    =
    \lambda x
    +\frac{g_3}{2}x^2
    +\frac{g_4}{6}x^3
    +\cdots .
    \label{eq:supp_force_expansion}
\end{equation}
For a symmetric displacement \(\delta>0\) in \(x\), the odd part
satisfies
\begin{equation}
    \frac{
        \mathcal N(\delta)-\mathcal N(-\delta)
    }{2\delta}
    =
    \lambda
    +\frac{g_4}{6}\delta^2
    +O(\delta^4).
    \label{eq:supp_odd_force}
\end{equation}
The numerical fit includes the \(\delta^4\) term.

The profile derivatives in
Eq.~\eqref{eq:supp_chi_u2} are obtained from a sixth-order symmetric
polynomial fit over
\begin{equation}
    \left|
        |\phi_c|-|\phi_c|_{\rm cusp}
    \right|
    \leq
    2\times10^{-4}.
    \label{eq:supp_profile_window}
\end{equation}
For force fits whose largest displacement \(\delta_{\max}\) lies in
\begin{equation}
    7\times10^{-4}
    \leq
    \delta_{\max}
    \leq
    1.6\times10^{-3},
\end{equation}
the extracted quartic coefficient is stable at
\begin{equation}
    g_4^{\rm LS}
    =
    -775.704231639 .
    \label{eq:supp_g4_LS}
\end{equation}
The fixed-charge condition is maintained to
\begin{equation}
    |Q(x)-Q_*|
    \sim10^{-15}.
\end{equation}

At the same resolution, the equilibrium-family identity gives
\begin{equation}
    g_4^{\rm id}
    =
    -775.770242609 ,
\end{equation}
so that
\begin{equation}
    \frac{
        |g_4^{\rm LS}-g_4^{\rm id}|
    }{
        |g_4^{\rm id}|
    }
    =
    8.51\times10^{-5}.
    \label{eq:supp_g4_LS_difference}
\end{equation}
Here ``LS'' denotes the off-shell Lyapunov--Schmidt estimate and
``id'' the equilibrium-family identity.  The off-shell result therefore
checks the quartic coefficient without using \(Q'''\).

Because the coordinate in Eq.~\eqref{eq:supp_x_normalisation} differs
from \(|\phi_c|-|\phi_c|_{\rm cusp}\) only by a quadratic near-identity
reparameterisation, the two quartic coefficients can be compared
directly at the endpoint: the corresponding shift is proportional to
\(g_3\), which vanishes there.

The profile-fit window cannot be reduced indefinitely.  At a
half-width \(5\times10^{-5}\), discretisation noise in the second
profile derivative becomes visible and the extracted \(g_4\) begins
to drift.  We therefore use the \(2\times10^{-4}\) symmetric window
for the quoted off-shell result.

\section{Critical soft-mode scaling}

To determine the near-endpoint spectral scaling, we evaluate the
fundamental radial mode from within the shrinking stable region.  For
each \(\sigma<\sigma_{\rm cusp}\), the mode is evaluated at the
midpoint of the two zero-mode boundaries,
\begin{equation}
    |\phi_c|_{\rm mid}
    =
    \frac{
        |\phi_c|_-+|\phi_c|_+
    }{2}.
    \label{eq:supp_midpoint}
\end{equation}
This follows the centre of the stable interval up to higher-order
corrections and avoids the boundaries themselves, where
\(\Omega_0^2=0\) by construction.  For each \(\sigma\), the same
midpoint is used in the three pulsation calculations so that the
resolution dependence of the spectrum can be compared directly.

At finite resolution, the refined cusp retains the small positive
residual listed in Table~\ref{tab:cusp_convergence}.  We subtract it
separately at each resolution,
\begin{equation}
    \Delta\Omega_0^2
    \left(
        N_{\rm int},\sigma
    \right)
    =
    \Omega_0^2
    \left(
        N_{\rm int},\sigma
    \right)
    -
    \Omega_{0,\rm cusp}^2
    \left(
        N_{\rm int}
    \right).
    \label{eq:supp_soft_subtraction}
\end{equation}
The distance from the endpoint is likewise measured relative to the
refined cusp at the same resolution,
\begin{equation}
    \Delta\sigma_N
    =
    \sigma_{{\rm cusp},N}-\sigma .
    \label{eq:supp_delta_sigma}
\end{equation}
Here \(N\) labels the corresponding radial resolution
\(N_{\rm int}\).  This removes the small finite-grid shifts of the
eigenvalue and cusp location from the scaling analysis.

\begin{figure}[t]
    \centering
    \includegraphics[width=\columnwidth]
    {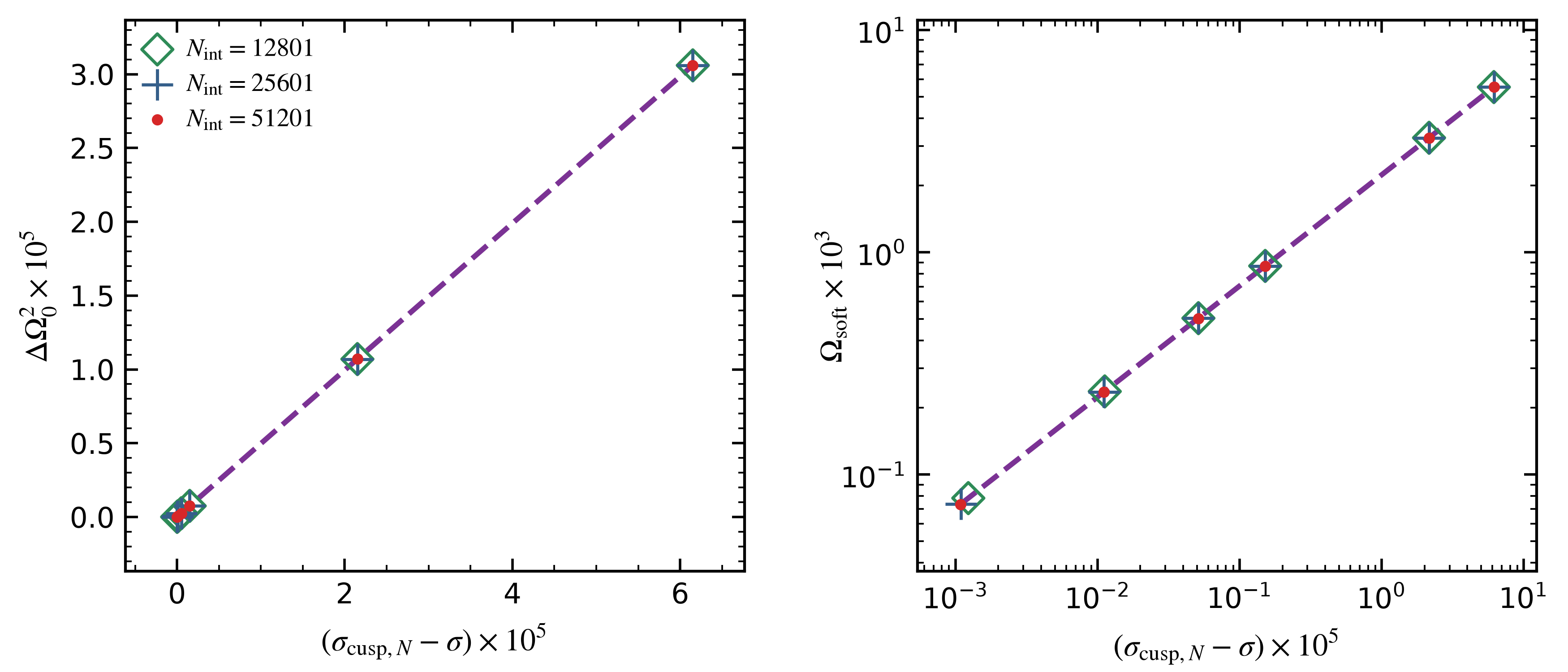}
    \caption{
    Critical softening of the fundamental radial mode.
    The left panel shows the residual-corrected squared eigenvalue
    \(\Delta\Omega_0^2\) against
    \(\Delta\sigma_N=\sigma_{{\rm cusp},N}-\sigma\) for
    \(N_{\rm int}=12801\), \(25601\), and \(51201\).
    The cusp residual and cusp location are treated separately at each
    resolution.  The three data sets nearly coincide and follow
    \(\Delta\Omega_0^2\simeq0.49746\,\Delta\sigma_N\).
    The right panel shows
    \(\Omega_{\rm soft}=\sqrt{\Delta\Omega_0^2}\) on logarithmic axes,
    with
    \(\Omega_{\rm soft}\propto\Delta\sigma_N^{1/2}\).
    }
    \label{fig:supp_soft_scaling}
\end{figure}

The left panel of Fig.~\ref{fig:supp_soft_scaling} shows that the
corrected squared eigenvalue vanishes linearly with
\(\Delta\sigma_N\).  Fits through the origin give
\begin{align}
    C_{12801}
    &=
    0.497398059464,
    \\
    C_{25601}
    &=
    0.497444584405,
    \\
    C_{51201}
    &=
    0.497456310847,
    \label{eq:supp_soft_slopes}
\end{align}
where
\begin{equation}
    \Delta\Omega_0^2
    =
    C_N\,\Delta\sigma_N .
    \label{eq:supp_soft_linear}
\end{equation}
The slopes agree closely across the three resolutions, and the
through-origin fits have \(R^2>0.99999999\).

As a check on the subtraction, we also fit the unshifted eigenvalues as
\begin{equation}
    \Omega_0^2
    =
    b_N+\widetilde C_N\Delta\sigma_N ,
    \label{eq:supp_soft_raw_fit}
\end{equation}
where \(b_N\) and \(\widetilde C_N\) are the free intercept and slope.
The fitted intercepts reproduce the independently measured
\(\Omega_{0,\rm cusp}^2(N_{\rm int})\) to within
\(10^{-10}\)--\(10^{-9}\).

The near-endpoint scaling may therefore be written as
\begin{equation}
    \Delta\Omega_0^2
    \simeq
    0.49746\,
    (\sigma_{\rm cusp}-\sigma).
    \label{eq:supp_soft_linear_limit}
\end{equation}
A direct power-law fit,
\begin{equation}
    \Delta\Omega_0^2
    \propto
    (\Delta\sigma_N)^p ,
    \label{eq:supp_soft_power}
\end{equation}
gives
\begin{equation}
    p
    =
    0.99994596,\quad
    1.00007571,\quad
    1.00003710
    \label{eq:supp_soft_exponents}
\end{equation}
for \(N_{\rm int}=12801\), \(25601\), and \(51201\), respectively.
The data are consistent with a linear vanishing of
\(\Delta\Omega_0^2\) at the endpoint.

For the scaling analysis we define the residual-corrected soft-mode
frequency
\begin{equation}
    \Omega_{\rm soft}
    \equiv
    \sqrt{\Delta\Omega_0^2}.
    \label{eq:supp_soft_frequency_def}
\end{equation}
Equation~\eqref{eq:supp_soft_linear_limit} then gives
\begin{equation}
    \Omega_{\rm soft}
    \simeq
    0.705306\,
    (\sigma_{\rm cusp}-\sigma)^{1/2}.
    \label{eq:supp_soft_frequency}
\end{equation}
Equivalently, fitting \(\Omega_{\rm soft}\) on logarithmic axes gives
the exponents
\begin{equation}
    0.49997298,\quad
    0.50003785,\quad
    0.50001855 ,
    \label{eq:supp_frequency_exponents}
\end{equation}
for increasing radial resolution.  The right panel of
Fig.~\ref{fig:supp_soft_scaling} shows the square-root behaviour.

At finite resolution, \(\Omega_{\rm soft}\) is defined after removing
the independently measured cusp residual.  Since
\(\Omega_{0,\rm cusp}^2\rightarrow0\) in the continuum limit,
\(\Omega_{\rm soft}\) coincides there with the fundamental radial
frequency near the endpoint.

The corresponding linear soft-mode timescale,
\begin{equation}
    \tau_{\rm soft}
    \equiv
    \Omega_{\rm soft}^{-1},
\end{equation}
therefore behaves as
\begin{equation}
    \tau_{\rm soft}
    \simeq
    1.417825\,
    (\sigma_{\rm cusp}-\sigma)^{-1/2}.
    \label{eq:supp_soft_timescale}
\end{equation}
This timescale is defined by the linear radial eigenfrequency and is
not a measurement of the nonlinear relaxation time.

\section{Robustness under potential deformation}

We also test the sensitivity of the endpoint to the scalar
self-interaction by deforming the sextic coefficient,
\begin{equation}
    V_\epsilon(|\phi|)
    =
    |\phi|^2
    -\frac{4|\phi|^4}{\sigma^2}
    +\frac{4(1+\epsilon)|\phi|^6}{\sigma^4},
    \qquad
    -0.10\leq\epsilon\leq0.10 ,
    \label{eq:supp_deformed_potential}
\end{equation}
so that \(\epsilon=0\) recovers the original solitonic potential.

For each fixed \(\epsilon\), we locate the cusp by solving
\begin{equation}
    Q'=0,
    \qquad
    Q''=0 ,
    \label{eq:supp_deformed_cusp_conditions}
\end{equation}
for \((|\phi_c|,\sigma)\), where the derivatives are taken at fixed
\(\sigma\) and \(\epsilon\).  We then evaluate
\(g_4=-\omega'Q'''\) and solve the radial pulsation problem at the
refined cusp.  The results at \(N_{\rm int}=25601\) are listed in
Table~\ref{tab:potential_deformation}.

\begin{table}[t]
    \centering
    \caption{
    Endpoint under a deformation of the sextic self-interaction, at
    \(N_{\rm int}=25601\).  The endpoint shifts across the sampled
    range while \(g_4\) remains finite and negative.  The final column
    gives the finite-grid fundamental radial eigenvalue at the refined
    cusp.
    }
    \label{tab:potential_deformation}
    \begin{ruledtabular}
    \begin{tabular}{ccccc}
        \(\epsilon\)
        & \(\sigma_{\rm cusp}\)
        & \(|\phi_c|_{\rm cusp}\)
        & \(g_4\)
        & \(\Omega_{0,\rm cusp}^2\)
        \\
        \hline
        \(-0.10\)
        & 0.3181298122
        & 0.1968271371
        & \(-744.8365\)
        & \(1.61868\times10^{-6}\)
        \\
        \(-0.05\)
        & 0.3195996361
        & 0.1932368787
        & \(-761.1062\)
        & \(1.54309\times10^{-6}\)
        \\
        \(0\)
        & 0.3210615109
        & 0.1898964835
        & \(-775.7713\)
        & \(1.47338\times10^{-6}\)
        \\
        \(+0.05\)
        & 0.3225146113
        & 0.1867775148
        & \(-788.9237\)
        & \(1.40898\times10^{-6}\)
        \\
        \(+0.10\)
        & 0.3239583397
        & 0.1838557755
        & \(-800.6393\)
        & \(1.34938\times10^{-6}\)
    \end{tabular}
    \end{ruledtabular}
\end{table}

The \(\epsilon=0\) entry comes from the deformation scan and agrees
with the dedicated undeformed refinement above; the small differences
in the final digits reflect the separate local fits.

The endpoint varies systematically across the sampled values of
\(\epsilon\).  Over the interval,
\begin{equation}
    \sigma_{\rm cusp}:
    \quad
    0.31813
    \longrightarrow
    0.32396 ,
\end{equation}
while
\begin{equation}
    |\phi_c|_{\rm cusp}:
    \quad
    0.19683
    \longrightarrow
    0.18386 .
\end{equation}
The quartic coefficient remains well separated from zero,
\begin{equation}
    -800.64
    \lesssim
    g_4
    \lesssim
    -744.84 .
    \label{eq:supp_deformed_g4_range}
\end{equation}
The quartic non-degeneracy of the undeformed endpoint therefore
persists throughout the tested deformation.

The radial calculation also identifies the nodeless fundamental mode
at every refined cusp.  The finite-resolution values in
Table~\ref{tab:potential_deformation} lie between
\(1.35\times10^{-6}\) and \(1.62\times10^{-6}\), comparable to the
undeformed residual at the same resolution.

The radial calculation is repeated at
\(N_{\rm int}=12801\) and \(25601\) for
\begin{equation}
    \epsilon=-0.05,\qquad 0,\qquad +0.05 .
\end{equation}
At each resolution the eigenvalue is evaluated at the corresponding
refined cusp.  On doubling the radial resolution, the residuals
decrease by factors
\begin{equation}
    3.99847,\qquad
    3.99846,\qquad
    3.99847 ,
    \label{eq:supp_deformed_radial_ratios}
\end{equation}
respectively.  These spot checks are consistent with the same
second-order decay established for the undeformed endpoint in
Table~\ref{tab:cusp_convergence}.  They support the interpretation of
the small positive values in Table~\ref{tab:potential_deformation} as
finite-grid residuals rather than physical gaps.

The endpoint, radial soft mode, and quartic non-degeneracy thus persist
over the tested \(\pm10\%\) variation of the sextic coefficient.
Within this family of deformations, the cusp-type endpoint is
numerically robust.  We do not infer robustness under arbitrary
changes of the scalar self-interaction.

\bibliography{BS}
\end{document}